# Reducing measurements in quantum erasure correction by quantum local recovery

**Ryutaroh Matsumoto**[1]

Received: 1 February 2026 / Revised: 1 February 2026 / Accepted: 22 July 2026

**Abstract**
As measurements are costly and prone to errors on certain quantum computing devices, we should reduce the number of measurements and the number of measured qudits as small as possible in quantum erasure correction. It is intuitively obvious that a decoder can omit measurements of stabilizers that are irrelevant to erased qudits, but this intuition has not been rigorously formalized as far as the author is aware. In this paper, we formalize relevant stabilizers sufficient to correct erased qudits with a quantum stabilizer code, by using a recent idea from quantum local recovery. The minimum required number of measured stabilizer observables is also clarified. As an application, we also show that correction of $\delta$ erasures on a generalized surface code proposed by Delfosse, Iyer and Poulin requires at most $\delta$ measurements of vertexes and at most $\delta$ measurements of faces, independently of its code parameters.

## 1 Introduction

An *erasure* in quantum and classical error correction means an error whose position in a codeword is known (Bennett et al. 1997; Grassl et al. 1997). It is known that a quantum error-correcting code can correct twice as many erasures as errors. In light of this, recent papers (Wu et al. 2022; Kang et al. 2023) take advantage of erasures in the quantum fault-tolerant computation, as some physical devices allow identification of qubits with erasures in a codeword without destruction of encoded quantum information or stabilizer measurements (Wu et al. 2022; Kang et al. 2023).

The stabilizer code (Gottesman 1996; Calderbank et al. 1997, 1998; Ashikhmin and Knill 2001; Ketkar et al. 2006) is a class of quantum error-correcting codes allowing efficient implementation of encoders and decoders, and in this paper we focus on erasure correction

✉ Ryutaroh Matsumoto
ryutaroh@ict.e.titech.ac.jp

1 Department of Information and Communications Engineering, Institute of Science Tokyo, Ookayama 2-12-1, Meguro, Tokyo 152-8550, Japan

with stabilizer codes, and its subclass, the generalized surface codes (Delfosse et al. 2016; Delfosse and Zémor 2020).

Although erasure correction for stabilizer codes can be efficiently done without measurements (Chiwaki and Matsumoto 2025), the standard and popular way (Delfosse and Zémor 2020) of erasure correction involves measurements. Measurements are costly on some physical devices and measurement-free fault-tolerant computation has been actively investigated recently (Perlin et al. 2023; Heußen et al. 2024; Veroni et al. 2024). It is also investigated to reduce the number of measurements in quantum *error* correction (Zhou et al. 2025). This kind of optimization seems unexplored in recent studies of quantum *erasure* correction (Delfosse and Zémor 2020; Wu et al. 2022; Kang et al. 2023; Connolly et al. 2024) (see Remark 16 for details), while the quantum local recovery (Galindo et al. 2026; Golowich and Guruswami 2025; Luo et al. 2025; Sharma et al. 2025; Xie et al. 2025; Galindo et al. 2025; Li et al. 2025, 2026; Bu et al. 2025; Cao and Zhou 2025; Zhou and Cao 2025) studies reducing the number of measured qudits without consideration of the number of measurements by a decoder. As stated in Theorem 5, the observation in this paper reduces difficulty of measurements. This paper proposes a method to reduce the number of measurements in quantum erasure correction.

We will mostly consider erasure-only correction, in the same way as Connolly et al. (2024) and Delfosse and Zémor (2020). In Connolly et al. (2024) and Delfosse and Zémor (2020), the authors proposed the novel erasure-only decoding algorithms and investigated their algebraic and algorithmic aspects, unlike more practical studies (Chadwick et al. 2025; Fujii and Tokunaga 2012; Stace et al. 2009; Nagayama et al. 2017) which proposed the error-and-erasure decoding algorithms for surfaces codes and investigated their more practical aspects, such as their thresholds of quantum fault tolerance. This paper shares the same spirit as Connolly et al. (2024) and Delfosse and Zémor (2020), will investigate algebraic and algorithmic aspects of the proposed algorithms, and leave investigation of their more practical aspects as a future research agenda. The error-and-erasure correction will be briefly considered in Remark 7.

In erasure correction of a classical linear code of length $n$ with a parity check matrix $H$ and a set $I \subset \{1, \ldots, n\}$ of erased symbols, we only need to consider rows in $H$ whose components in $I$ are not all zeros, and one can just ignore rows in $H$ whose components in $I$ are all zeros. The essential idea in this paper is to carry over this observation for classical linear codes to erasure correction with quantum stabilizer codes as stated in Proposition 4.

After reviewing quantum stabilizer codes in Sect. 2, we will clarify which stabilizer measurements are sufficient for correction of erased qudits at positions $I$ in Theorem 3 and Proposition 4, which lead to clarification of the minimum required number of measurements by our proposal in Theorems 5 and 8, in Sect. 3. In Sect. 4, we specialize results in Sect. 3 to the CSS codes as a preparation of studying generalized surface codes (Delfosse et al. 2016; Delfosse and Zémor 2020). As a consequence of Sect. 4, in Sect. 5, we will clarify which stabilizer measurements are sufficient on which qubits, in order to correct given erasures on a generalized surface code. Examples in Sect. 6 illustrate results in Sect. 5. Concluding remarks are given in Sect. 7.

## 2 A review of *p*-ary quantum stabilizer codes

Let $p$ be a prime integer. Denote by $\mathcal{H}_p$ the $p$-dimensional complex linear spaces. A quantum state in $\mathcal{H}_p$ is called a *qudit*. When quantum messages and codewords are quantum states in tensor products of $p^m$-dimensional complex linear spaces, they can also be regarded as tensor products of $\mathcal{H}_p$. So we may assume without much loss of generality that each quantum symbol in messages and codewords belongs to $\mathcal{H}_p$, and we will consider stabilizer codes whose codewords belong to $\mathcal{H}_p^{\otimes n}$.

Let $\mathbf{F}_p = \{0, 1, \dots, p-1\}$ be the finite field with $p$ elements. We will use the notation of vectors and their associated unitary matrices introduced by Calderbank et al. (1998), which is one of the standard notations in expression of quantum stabilizer codes by finite field vectors. For a vector $\vec{x} = (a_1, \dots, a_n | b_1, \dots, b_n) \in \mathbf{F}_p^{2n}$ and $J \subset \{1, \dots, n\}$, let $P_J(\vec{x})$ be the projection map from $\mathbf{F}_p^{2n}$ to $\mathbf{F}_p^{2|J|}$, sending $\vec{x}$ to $(a_{i_1}, \dots, a_{i_{|J|}} | b_{i_1}, \dots, b_{i_{|J|}})$, where $\{i_1, \dots, i_{|J|}\} = J$. By $\mathbf{F}_p^J \subset \mathbf{F}_p^{2n}$ we mean $\{(a_1, \dots, a_n | b_1, \dots, b_n) \in \mathbf{F}_p^{2n} : a_i = b_i = 0 \text{ if } i \notin J\}$.

For a vector $\vec{x} = (a_1, \dots, a_n | b_1, \dots, b_n) \in \mathbf{F}_p^{2n}$, we define the associated $p^n \times p^n$ unitary matrix $M(\vec{x})$ as $X^{a_1} Z^{b_1} \otimes \cdots \otimes X^{a_n} Z^{b_n}$, where $X|i\rangle = |1 + i \bmod p\rangle$ and $Z|i\rangle = \exp(2\pi\sqrt{-1}/p)^i |i\rangle$. We define $\omega = \exp(2\pi\sqrt{-1}/p)$ for $p \geq 3$ and $\omega = \sqrt{-1}$ for $p = 2$. We will consider the error group $E_n = \{\omega^i X^{a_1} Z^{b_1} \otimes \cdots \otimes X^{a_n} Z^{b_n} : a_1, b_1, \dots, a_n, b_n \in \mathbf{F}_p$ and $i$ is an integer $\}$.

We always consider the standard symplectic inner product $\langle \cdot, \cdot \rangle$ in $\mathbf{F}_p^{2n}$, that is, the symplectic inner product of $(a_1, \dots, a_n | b_1, \dots, b_n)$ and $(a'_1, \dots, a'_n | b'_1, \dots, b'_n)$ is $\sum_{i=1}^n (a_i b'_i - a'_i b_i)$, and $C^\perp$ denotes the orthogonal space of $C$ in $\mathbf{F}_p^{2n}$ with respect to that inner product. Let $S \subset E_n$ be a commutative subgroup. There exists a common eigenvector $|\varphi\rangle$ such that, for any $U \in S$, $U|\varphi\rangle = \eta(U)|\varphi\rangle$, where $\eta(U)$, not necessarily equal to $+1$ but often chosen to be $+1$, is the eigenvalue of $|\varphi\rangle$ to $U$. Observe that $\eta(U)$ is always a power of $\omega$. The quantum stabilizer code $Q(S) \subset \mathcal{H}_p^{\otimes n}$ with parameter $[[n, k]]_p$ is defined as $\{|\psi\rangle \in \mathcal{H}_p^{\otimes n} : U|\psi\rangle = \eta(U)|\psi\rangle$ for all $U \in S\}$ (Calderbank et al. 1997, 1998; Ashikhmin and Knill 2001). The normalizer $S'$ of $S$ in $E_n$ is $S' = \{V \in E_n : UV = VU$ for all $U \in S\}$.

Let $C = \{\vec{x} \in \mathbf{F}_p^{2n} : \omega^i M(\vec{x}) \in S$ for some integer $i\}$. Then $C$ is an $\mathbf{F}_p$-linear space with $\dim C = n - k$, $C^\perp = M^{-1}(S')$, $C^\perp \supset C$, and $\dim C^\perp = n + k$. By abuse of notation, we also write $Q$(the commutative subgroup generated by $\{M(\vec{x}) : \vec{x} \in C\}$) as $Q(C)$. Since $M(\vec{x})$ may not have $+1$ eigenvalue for $p = 2$, we do not require $\eta(M(\vec{x}))$ to be $+1$ in this paper. For example, there is no eigenvector with eigenvalue $+1$ when one-dimensional $C$ is spanned by $(1, 1, 1|1, 1, 1) \in \mathbf{F}_2^6$. The requirement of eigenvalue being $+1$ prevents us from using some self-orthogonal spaces $C \subset \mathbf{F}_2^{2n}$ for defining stabilizer codes $Q(C)$. Ball, Moreno and Simoens precisely stated the subtlety and the prescription of defining $Q(C)$ with eigenvalue $+1$ from $C \subset \mathbf{F}_2^{2n}$ (Ball et al. 2025, Section II.D).

**Assumption 1** *(The number of measurements by a decoder)* Suppose that a self-orthogonal linear space $C \subset \mathbf{F}_p^{2n}$ defines a quantum stabilizer code $Q(C)$, and $\{\vec{x}_1, \dots, \vec{x}_{n-k}\}$ is a basis of $C$. A decoder defined by $C$ is assumed to measure each observable whose eigenspaces are the same as $M(\vec{x}_i)$ for $i = 1, \dots, n-k$, and the total number of measurements is $n - k = \dim C$.

***Remark 2*** When the dimension of a qudit is not a prime $p$ but a prime power $p^m$, the number of measurements might not be equal to $\dim C$ as in Assumption 1. By considering the relation between $p^m$-ary quantum codes and self-orthogonal linear spaces $C \subset \mathbf{F}_{p^m}^{2n}$ shown in Ashikhmin and Knill (2001), it still seems reasonable to assume that the number of measurements is *proportional to* $\dim C$. All the mathematical claims in this paper remain valid even when the size of the finite field is a prime power $p^m$, and the results presented here are also useful for reducing the number of measurements for $p^m$-ary codes.

## 3 Localized correction of erasures with a stabilizer code

Suppose that we have erasures at $I \subset \{1, \ldots, n\}$. We assume that $Q(C)$ can correct erasures at $I$, which is equivalent to Galindo et al. (2026)

$$C \cap \mathbf{F}_p^I = C^{\perp} \cap \mathbf{F}_p^I. \tag{1}$$

Decompose $C$ into

$$C = D + (C \cap \mathbf{F}_p^{\overline{I}}) \tag{2}$$

as the sum of linear spaces $D$ and $C \cap \mathbf{F}_p^{\overline{I}}$. **We propose to correct erasures at $I$ on a codeword in $Q(C)$ by measuring observables defined by basis vectors of $D$.** Observe that stabilizer (complex) matrices corresponding to $C \cap \mathbf{F}_p^{\overline{I}}$ via the mapping $M(\cdot)$ have the identity matrices at $I$, and seem of no use for correcting erasures at $I$. We will show that it is sufficient for a decoder to measure observables corresponding to basis vectors of $D$. In the standard decoding we make $n - k = \dim C$ times of measurements by Assumption 1, but the number of measurements will be reduced to $\dim D$.

For $\vec{x} \in \mathbf{F}_p^{2n}$ let $\mathrm{supp}(\vec{x}) = \{1 \le i \le n\colon \vec{x}$ has nonzero component at index $i$ or $n+i\}$, and $\mathrm{supp}(D) = \bigcup_{\vec{x} \in D} \mathrm{supp}(\vec{x})$. We see that $\mathrm{supp}(D) \subseteq \{1, \ldots, n\}$. In the quantum local recovery (Galindo et al. 2026; Golowich and Guruswami 2025; Luo et al. 2025; Sharma et al. 2025; Xie et al. 2025; Galindo et al. 2025; Li et al. 2026, 2025; Bu et al. 2025; Cao and Zhou 2025; Zhou and Cao 2025), when erasures at $I$ are corrected, a set $J \subset \{1, \ldots, n\}$ of qudits in a partially erased quantum codeword is used to recover qudits at the erased positions $I$, where $J \supset I$. We will call $J$ a *recovering set*. As $D \subset C \cap \mathbf{F}_p^{\mathrm{supp}(D)}$, our proposal is actually local recovery of erasures at $I$ by a recovering set $\mathrm{supp}(D)$. Since $D \subset C$, any stabilizer codeword in $Q(C)$ also belongs to $Q(D)$ (with suitable choices of eigenvalues $\eta(M(\vec{x}))$ for $\vec{x} \in D$ in defining $Q(D)$).

**Theorem 3** *Let $D$ be an arbitrary subspace of $C$. Measurements of observables defined by $D$ can correct erasures at $I$ on codewords in $Q(C)$ if and only if*

$$C \cap \mathbf{F}_p^I = D^{\perp} \cap \mathbf{F}_p^I. \tag{3}$$

***Proof*** The proof below follows the argument in Galindo et al. (2026, pp. 10–11). Let $\vec{x}_1, \ldots, \vec{x}_\ell$ be a basis of $D$, and $\vec{e} \in \mathbf{F}_p^I$ be an erasure vector. Since any $M(\vec{e})$ for $\vec{e} \in C$ has no effect on codewords in $Q(C)$, a decoder only has to identify $\vec{e}$ up to modulo $C$. Each outcome of measuring an observable whose eigenspaces are the same as $M(\vec{x}_i)$ gives $s_i \in \mathbf{F}_p$ for $i = 1, \ldots, \ell$. By the same argument as Galindo et al. (2026), the system of linear equations $s_i = \langle \vec{x}_i, \vec{e} \rangle$ for $i = 1, \ldots, \ell$ gives a unique solution $\vec{e} \in \mathbf{F}_p^I$ modulo $C$ if and only if (3) holds. Therefore the theorem is proved. □

**Proposition 4** *Let $D$ be any subspace of $C$. Then (3) implies both (1) and (2). On the other hand, (1) and (2) imply (3), which means that measurements of observables defined by $D$ can correct erasures at $I$ on codewords in $Q(C)$ when both (1) and (2) hold.*

***Proof*** By Galindo et al. (2019, Section 4.1), (3) is equivalent to

$$P_I(C^{\perp}) = P_I(D). \tag{4}$$

Suppose that (2) is false, that is, there exists $\vec{x} \in C \setminus (D + (C \cap \mathbf{F}_p^{\overline{I}}))$. Since $\vec{x} \notin D + (C \cap \mathbf{F}_p^{\overline{I}})$, for any $\vec{y} \in D$ we have $\vec{x} - \vec{y} \notin C \cap \mathbf{F}_p^{\overline{I}}$. Since $\ker(P_I) = \mathbf{F}_p^{\overline{I}}$, we have $P_I(\vec{x}) \notin P_I(D)$ and

(4) is false as $C \subset C^{\perp}$. We have shown the contraposition of implication (3) $\Rightarrow$ (2). Since $D^{\perp} \supset C^{\perp}$, we also see (3) $\Rightarrow$ (1).

We will show the reverse implication. Suppose that both (1) and (2) are true. Since $\ker(P_I) = \mathbf{F}_p^{\overline{I}}$, by (2) we have $P_I(C) = P_I(D)$. By Galindo et al. (2019, Section 4.1), (1) is equivalent to $P_I(C) = P_I(C^{\perp})$, which implies (4) and equivalently (3). □

**Theorem 5** *By Proposition 4, the minimum number of measurements to correct erasures at $I$ by our proposal is*

$$\min_{D \subseteq C} \{\dim D : C = D + (C \cap \mathbf{F}_p^{\overline{I}})\} = \dim C - \dim C \cap \mathbf{F}_p^{\overline{I}}, \tag{5}$$

*which is*

$$\leq 2|I|. \tag{6}$$

*The minimum in (5) is attained if and only if*

$$D \cap \mathbf{F}_p^{\overline{I}} = \{\mathbf{0}\}. \tag{7}$$

*The worst-case number of measurements to correct $\delta$ erasures by our proposal is*

$$\dim C - \min_{I \subset \{1,\ldots,n\}, |I| = \delta} \dim C \cap \mathbf{F}_p^{\overline{I}}. \tag{8}$$

***Proof*** Claims (5) and (8) are obvious. We see that

$$\dim D = \dim C - \dim C \cap \mathbf{F}_p^{\overline{I}} \tag{9}$$

holds if and only if we choose $D$ such that

$$D \cap (C \cap \mathbf{F}_p^{\overline{I}}) = \emptyset. \tag{10}$$

Equality (10) holds if and only if (7) does because $D \subset C$.

Since $C \cap \mathbf{F}_p^{\overline{I}} = C \cap \ker(P_I)$, we have

$$\begin{aligned} \dim C - \dim C \cap \mathbf{F}_p^{\overline{I}} &= \dim P_I(C) \\ &\leq 2|I|, \end{aligned}$$

implying (6). □

***Remark 6*** Let $\vec{x}_1, \ldots, \vec{x}_{n-k}$ be row vectors in a reduced row echelon form of a matrix whose row space is $C$, such that for each index $i \in I$ there is at most one vector $\vec{x}_j$ whose $i$-th component is nonzero and at most one vector $\vec{x}_m$ whose $(i+n)$-th component is nonzero. Such a reduced row echelon form is obtained by application of elementary row operations after reordering of column indices that moves $I \cup \{i + n : i \in I\}$ to $\{1, \ldots, 2|I|\}$. A basis of $D$ with $D \cap \mathbf{F}_p^{\overline{I}} = \{\mathbf{0}\}$ can be found as $\{\vec{x}_i : \mathrm{supp}(\vec{x}_i) \cap I \neq \emptyset\}$, and that of $C \cap \mathbf{F}_p^{\overline{I}}$ can be found as $\{\vec{x}_i : \mathrm{supp}(\vec{x}_i) \cap I = \emptyset\}$. Note that the computational complexity of finding $\vec{x}_1, \ldots, \vec{x}_{n-k}$ is cubic in $n$ as it is Gaussian elimination performed once on an $(n-k) \times 2n$ matrix.

***Remark 7*** Observe that $C \cap \mathbf{F}_p^{\overline{I}}$ is also self-orthogonal and defines a stabilizer code of length $n - |I|$ with the minimum distance, say, $d^{\overline{I}}$. In practice, it is uncertain whether or not there also exists an error in $\overline{I}$. We can ensure that the number of errors at $\overline{I}$ is either zero or $\geq d^{\overline{I}}$ by measuring the stabilizer generators defined by $C \cap \mathbf{F}_p^{\overline{I}}$, and we retain the advantage of reduced measurements.

Theorem 5 shows the minimum required numbers of measurements by our proposal. The next theorem will show different expressions of the minimum numbers given in Theorem 5. Remark 10 will discuss relation between Theorems 5 and 8.

**Theorem 8** *Let $\vec{x}_1, \ldots, \vec{x}_\ell \in C$, and $D \subseteq C$ be the linear space spanned by $\vec{x}_1, \ldots, \vec{x}_\ell$. If a decoder for $Q(C)$ can correct erasures at $I$ by measuring observables corresponding to $\vec{x}_1, \ldots, \vec{x}_\ell$, then (3) holds. Therefore, the minimum number of measurements for correcting erasures at $I$ on codewords in $Q(C)$ is*

$$\min_{D \subseteq C} \{\dim D : C \cap \mathbf{F}_p^I = D^\perp \cap \mathbf{F}_p^I\}. \tag{11}$$

*For a vector $\vec{x} \in \mathbf{F}_p^{2n}$, let $w_s(\vec{x}) = |\mathrm{supp}(\vec{x})|$, the symplectic weight of $\vec{x}$. Then the minimum cardinality of a fixed set of observables for correcting any $\delta$ erasures on codewords in $Q(C)$ is*

$$\min_{D \subseteq C} \{\dim D : w_s(D^\perp \setminus C) \geq \delta + 1\} \tag{12}$$

$$= 2n - \max_{D^\perp \supset C^\perp} \{\dim D^\perp : w_s(D^\perp \setminus C) \geq \delta + 1\}, \tag{13}$$

*where $w_s(D^\perp \setminus C) = \min\{w_s(\vec{x}) : \vec{x} \in D^\perp \setminus C\}$, which is greater than or equal to the minimum distance $w_s(D^\perp \setminus D)$ of the quantum stabilizer code $Q(D)$.*

***Proof*** By Theorem 3, (3) is a necessary and sufficient condition for correction of erasures at $I$ on codewords in $Q(C)$ by measuring observables corresponding to $\vec{x}_1, \ldots, \vec{x}_\ell$.

Observe that (3) is equivalent to

$$(D^\perp \cap \mathbf{F}_p^I) \setminus (C \cap \mathbf{F}_p^I) = \emptyset. \tag{14}$$

If (14) holds for every $I$ with $|I| = \delta$ then we have

$$w_s(D^\perp \setminus C) \geq \delta + 1. \tag{15}$$

We see that the minimum number of measurements for correcting any $\delta$ erasures is greater than or equal to (12).

On the other hand, any $D \subset C$ with (15) satisfies (3) and (14). So a linear subspace $D \subset C$ whose dimension attains the minimum (12) while satisfying (15) can correct any $\delta$ or less erasures by measuring observable corresponding to basis vectors of $D$, completing the proof. □

***Remark 9*** The combinatorial quantities (8), (12) and (13) have engineering significance in quantum erasure correction, but we are not aware of their relations to known combinatorial quantities of linear codes or quantum codes, such as the generalized Hamming weight (Wei 1991; Helleseth et al. 1992), the relative generalized Hamming weight (Luo et al. 2005), the dimension length profile (Forney 1994), or the relative generalized symplectic weight (Matsumoto 2020). Investigation of (8), (12) and (13) might be an interesting future research agenda. We note that (12) and (13) look similar to the dimension length profile of a classical linear code (Forney 1994) that characterizes its trellis complexity.

***Remark 10*** For a given $C \subset \mathbf{F}_p^{2n}$ and $I \subset \{1, \ldots, n\}$, in light of Proposition 4, (5) and (11) are the same if (1) holds, while (11) is undefined if (1) is false.

On the other hand, the minimization in (8) allows choices of different subspaces $D \subset C$ depending on different sets $I$ of erasures, while (12) and (13) use only one subspace $D \subset C$ for every erasure set $I$ of size $\delta$. Therefore, (8) is generally smaller than (12) and (13).

## 4 Localized erasure correction with two-code CSS codes

Let $C_i$ be an $[n, k_i]$ classical linear code over $\mathbf{F}_p$, for $i = X, Z$, and $C_i^{\perp e}$ the Euclidean dual of $C_i$. Assume $C_Z \times C_X \subset C_X^{\perp e} \times C_Z^{\perp e}$. By using $C_Z \times C_X$ in place of $C$ in Sect. 3, we can define a $p$-ary $[[n, n - k_X - k_Z]]$ stabilizer code. Also observe that the measurements defined by basis vectors of $C_X$ identify $X$ erasures, and that the measurements defined by those of $C_Z$ identify $Z$ erasures.

For a vector $\vec{x} \in \mathbf{F}_p^n$ and a linear space $D \subseteq \mathbf{F}_p^n$, hereafter let $\mathrm{supp}(\vec{x}) = \{1 \le i \le n\colon \vec{x}$ has nonzero component at index $i\}$, $\mathrm{supp}(D) = \bigcup_{\vec{x} \in D} \mathrm{supp}(\vec{x})$ and $\mathbf{F}_p^I = \{(a_1, \ldots, a_n) \in \mathbf{F}_p^n : a_i = 0 \text{ if } i \notin I\}$, by abuse of notation. In a similar way to the argument in Sect. 2, define $D_i$ by $C_i = D_i + (C_i \cap \mathbf{F}_p^{\overline{I}})$ for $i = X, Z$. Then we have (by a proof similar to that of Proposition 4):

**Proposition 11** *The $X$ erasures can be corrected by* $\dim D_X$ *measurements of qudits in* $\mathrm{supp}(D_X)$ *and a unitary matrix acting on those in* $\mathrm{supp}(D_X)$. *Similarly, the $Z$ erasures can be corrected by* $\dim D_Z$ *measurements of qudits in* $\mathrm{supp}(D_Z)$ *and a unitary matrix acting on those in* $\mathrm{supp}(D_Z)$. □

***Remark 12*** Similarly to Remark 6, let $\vec{x}_1, \ldots, \vec{x}_{k_X}$ be row vectors in the reduced row echelon form of a matrix whose row space is $C_X$, such that for each index $i \in I$ there is at most one vector $\vec{x}_j$ whose $i$-th component is nonzero. A basis of $D_X$ with $D_X \cap \mathbf{F}_p^{\overline{I}} = \{\mathbf{0}\}$ can be found as $\{\vec{x}_i\colon \mathrm{supp}(\vec{x}_i) \cap I \ne \emptyset\}$, and that of $C_X \cap \mathbf{F}_p^{\overline{I}}$ can be found as $\{\vec{x}_i\colon \mathrm{supp}(\vec{x}_i) \cap I = \emptyset\}$. Bases for $D_Z$ and $C_Z \cap \mathbf{F}_p^{\overline{I}}$ can be found in a similar way. Similarly to Theorem 5, we also have $\dim D_X \le |I|$ and $\dim D_Z \le |I|$.

## 5 Generalized surface codes

### 5.1 Review of generalized surface codes

We follow a definition of the generalized surface codes (Delfosse et al. 2016; Delfosse and Zémor 2020). A concrete example of a small surface code will be given in the next section. A topological surface with boundaries (Muñoz et al. 2020, Chapter 1) is denoted by a triple $(V, E, F)$, where $V$ is the vertex set, $E \subset V \times V$ is the edge set, and $F \subset 2^E$ is the face (or region) set. A face $f \in F$ is surrounded by edges $e \in f$, and an edge $e = \{u, v\}$ connects two vertexes $u$ and $v$, where pairs $(u, v)$ and $(v, u) \in V \times V$ are identified with a set $\{u, v\}$.

An edge $e$ is said to be a boundary if $|\{f \in F\colon e \in f\}| = 1$. A face $f$ is said to be boundary if $f$ contains a boundary edge. A vertex $v$ is boundary if there exists a boundary edge $e$ with $v \in e$.

Boundary edges, faces and vertex are categorized into open ones and closed ones. Each boundary edge is arbitrarily defined to be either open or closed. A boundary face $f$ is said to be open if $f$ contains an open boundary edge. A boundary vertex $v$ is said to be open if $v$ belongs to an open boundary edge. A boundary element that is not open is said to be closed. Denote by $\mathring{V}$ the set of non-boundary or closed boundary vertexes, and by $\mathring{E}$ the set of non-boundary or closed boundary edges. Then the generalized surface code has physical qubits corresponding to edges in $\mathring{E}$ and its stabilizer group is generated by $X_v$ for $v \in \mathring{V}$ and $Z_f$ for $f \in F$, where

$$X_e = \text{the Pauli } X \text{ matrix acting on qubit } e,$$

$$Z_e = \text{the Pauli } Z \text{ matrix acting on qubit } e,$$
$$X_v = \bigotimes_{v \in e} X_e,$$
$$Z_f = \bigotimes_{e \in f} Z_e.$$

## 5.2 Reduced measurements of at most 4|*I*| observables

With the above generalized surface code, a set $I$ of erasures can be identified with a subset of $\mathring{E}$. We will express $D_X$ and $D_Z$ in Sect. 4 in the language of the generalized surface codes.

**Proposition 13** *Let $\mathring{V}_{\overline{I}} = \{v \in \mathring{V}$: there exists no $e \in I$ such that $v \in e\}$, and $F_{\overline{I}} = \{f \in F$: there exists no $e \in I$ such that $e \in f\}$. Let $n = |\mathring{E}|$, then each edge $e \in \mathring{E}$ can be identified with an integer $i \in \{1, \ldots, n\}$. For $v \in \mathring{V}$, by $\vec{v} \in \mathbf{F}_2^n$ we denote the vector whose component is 1 if $v \in e$ and 0 otherwise, and for $f \in F$, by $\vec{f} \in \mathbf{F}_2^n$ we denote the vector whose component is 1 if $e \in f$ and 0 otherwise. Let $C_X$ be the linear code generated by $\{\vec{f}: f \in F\}$ and $C_Z$ be that generated by $\{\vec{v}: v \in \mathring{V}\}$. Then we have*

1. *The linear space generated by $\{\vec{f}: f \in F_{\overline{I}}\}$ is contained in $C_X \cap \mathbf{F}_2^{\overline{I}}$, and that generated by $\{\vec{v}: v \in \mathring{V}_{\overline{I}}\}$ is contained in $C_Z \cap \mathbf{F}_2^{\overline{I}}$.*
2. *The linear space generated by $\{\vec{f}: f \in F \setminus F_{\overline{I}}\}$ can be used as $D_X$ in Proposition 11, and that generated by $\{\vec{v}: v \in \mathring{V} \setminus \mathring{V}_{\overline{I}}\}$ can be used as $D_Z$ in Proposition 11.*
3. *The $X$ erasures can be identified by $|F \setminus F_{\overline{I}}|$ measurements of $|\bigcup_{f \in F \setminus F_{\overline{I}}} f|$ qubits, and the $Z$ erasures can be identified by $|\mathring{V} \setminus \mathring{V}_{\overline{I}}|$ measurements of $|\{e \in \mathring{E}$: there exists $v \in \mathring{V} \setminus \mathring{V}_{\overline{I}}$ such that $v \in e\}|$ qubits.*

***Proof*** Claim 1 immediately follows from the definitions. Claim 2 follows from Claim 1 and Proposition 11. Claim 3 follows from Proposition 11 and

$$\mathrm{supp}(D_X) = \bigcup_{f \in F \setminus F_{\overline{I}}} f,$$
$$\mathrm{supp}(D_Z) = \{e \in \mathring{E} : \exists v \in \mathring{V} \setminus \mathring{V}_{\overline{I}} \text{ such that } v \in e\}.$$

where $D_X$ and $D_Z$ are as given in Claim 2. □

Let $\mathring{V}_I = \{v \in \mathring{V}$: there exists *some* $e \in I$ such that $v \in e\}$, and $F_I = \{f \in F$: there exists *some* $e \in I$ such that $e \in f\}$. We have $\mathring{V}_I = \mathring{V} \setminus \mathring{V}_{\overline{I}}$ and $F_I = F \setminus F_{\overline{I}}$.

**Corollary 14** *The linear space generated by $\{\vec{f}: f \in F_I\}$ can be used as $D_X$ in Proposition 11, and that generated by $\{\vec{v}: v \in \mathring{V}_I\}$ can be used as $D_Z$ in Proposition 11. We have*

$$\dim D_X \le |F_I| \le 2|I|,$$
$$\dim D_Z \le |\mathring{V}_I| \le 2|I|.$$

***Proof*** Since $(V, E, F)$ is not a general undirected graph but a tiling on a surface, for each edge $e$, there are at most two vertexes and two faces adjacent to $e$, which implies $|F_I| \le 2|I|$ and $|\mathring{V}_I| \le 2|I|$. The rest of claims follows easily from the above. □

***Remark 15*** The total number of measurements is at most $4|I|$ to correct erasures at $I$ in our method in Sect. 5.2, which is larger than the value $2|I|$ given in Theorem 5. This difference in those numbers comes from the fact $(D_Z \times D_X) \cap \mathbf{F}_p^{\overline{I}} \neq \{\mathbf{0}\}$, which was required in Theorem 5. In Sect. 5.3 we show a procedure to compute fewer observables to identify erasures at $I$ with computational complexity $O(n^3)$.

***Remark 16*** Algorithms 1 and 2 in Delfosse and Zémor (2020) need outcomes from only the measurements clarified in Claim 3 in Proposition 13. However, the paper (Delfosse and Zémor 2020) did not explicitly state that the measurements not given in Claim 3 in Proposition 13 are unnecessary.

Note also that Algorithms 2 and 3 in a more recent paper (Connolly et al. 2024) on erasure correction also assume $(\dim C_X + \dim C_Z)$ stabilizer measurement outcomes for a CSS code defined by $C_Z \times C_X$, without explicitly stating which measurements are sufficient among $(\dim C_X + \dim C_Z)$ stabilizer measurements.

As a corollary to Proposition 13 and Corollary 14, we will clarify the locality of generalized surface codes. A quantum code is said to have locality $(r, \delta)$ if for any erasure at position $i$, there exists a local recovery set $J \ni i$ of size at most $r + \delta - 1$ that can locally recover any $\delta - 1$ erasures among qudits in $J$ (Galindo et al. 2026), and it is said to have locality $r$ if it has locality $(r, 2)$ (Golowich and Guruswami 2025). Recall that $E \subset V \times V$ and $F \subset E$. We emphasize that $\delta - 1$ erasures are considered in locality $(r, \delta)$.

**Corollary 17** *The generalized surface code considered in Sect. 5 has locality $(r, \delta)$ with*

$$r = \max_{\substack{I \subset \mathring{E}, \\ |I| = \delta}} \left| \{e' \in \mathring{E} : \exists e \in I, e \cap e' \neq \emptyset\} \cup \bigcup_{f \in F, I \cap f \neq \emptyset} f \right| - \delta + 1. \tag{16}$$

***Proof*** Equation (16) follows from the fact that edges in (16) are the set of edges adjacent to $\mathring{V}_I$ or $F_I$ in Corollary 14. □

For single erasure ($\delta = 2$ in locality $(r, \delta)$), estimation of the locality can be improved as follows. An example of Proposition 18 will be given in Sect. 6.2.

**Proposition 18** *The generalized surface code considered in Sect. 5 has locality $r$ with*

$$r = \max_{e \in \mathring{E}} \min_{v \in e, e \in f} \left| \{e' \in \mathring{E} : v \in e'\} \cup f \right| - 1. \tag{17}$$

***Proof*** Suppose that single erasure happened at an edge $e \in \mathring{E}$. Since the number of erasures is at most 1, identification of the erasure needs measurements of a single vertex $v \in e$ and single face $f$ containing $e$, with which qubits in $\{e' \in \mathring{E} : v \in e'\} \cup f$ are measured, which implies (17). □

### 5.3 Reduced measurements of at most $2|I|$ observables

In Sect. 5.2, we proposed easily computable sets $\mathring{V}_I$ of vertexes and $F_I$ of faces that can identify erasures at $I$, but their sizes are larger than the sufficient values given in Theorem 5. By using the procedure in Remark 12, one can reduce the number of observables to be measured, but resulting observables may not correspond to any vertex or face, which loses ease of implementation in surface codes. Below we propose a procedure to find a set of vertexes minimizing $\dim D_Z$ and the number of measurements to identify $Z$ erasures.

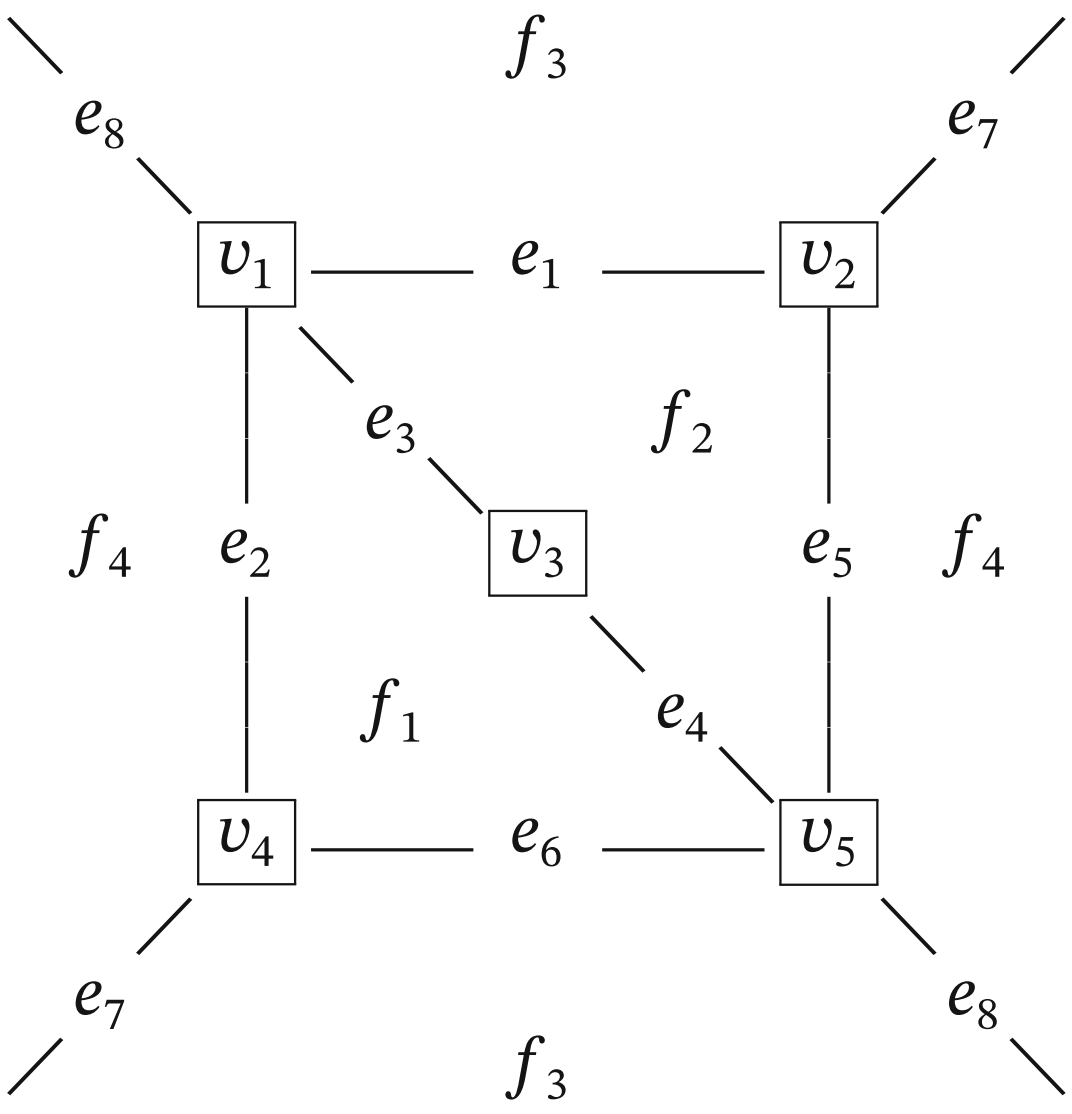

**Fig. 1** A [[8, 1, 2]] surface code obtained by removing an edge from Freedman and Meyer (2001, Fig. 2)

Let $\ell = \dim C_Z - \dim C_Z \cap \mathbf{F}_p^{\overline{I}}$. Let $\mathring{V} = \{v_1, \ldots, v_{k_Z}\}$, where $k_Z = |\mathring{V}|$. Build a $k_Z \times n$ matrix $A$ whose rows are $\{\vec{v}_i : v_i \in \mathring{V}\}$. Reorder column indices of $A$ so that indices in $I$ are moved to $\{1, \ldots, |I|\}$, perform elementary row operations on the column-permuted version of $A$, and obtain its row echelon form $B$. There are exactly $\ell$ rows $\vec{b}_{i_1}, \ldots, \vec{b}_{i_\ell}$ of $B$ that have at least one nonzero component among their leftmost $|I|$ components. Observe that $C_Z \cap \mathbf{F}_p^{\overline{I}}$ is spanned by $\{\vec{v}_i : i \notin \{i_1, \ldots, i_\ell\}\}$ and that a linear space $D_Z \subset C_Z$ with $D_Z \cap \mathbf{F}_p^{\overline{I}} = \{\mathbf{0}\}$ is spanned by $\{\vec{v}_{i_1}, \ldots, \vec{v}_{i_\ell}\}$, which are needed in Remark 12 to realize the minimum number of measurements. As stated in Remark 12, we have $\ell \le |I|$. A set of faces minimizing $\dim D_X$ can be found in a similar way. Since the procedure is Gaussian elimination, its computational complexity is cubic in $n$.

## 6 Examples with a small surface code

Fig. 1 shows a [[8, 1, 2]] surface code on a real projective plane with no boundary edge, obtained by removing an edge from the surface code given in Freedman and Meyer (2001, Fig. 2). With the notations in Sect. 5, we have $E = \mathring{E} = \{e_1, \ldots, e_8\}$, $V = \mathring{V} = \{v_1, \ldots, v_5\}$, and $F = \mathring{F} = \{f_1, \ldots, f_4\}$. Since any single error can be detected, its minimum distance is $\ge 2$. Because two errors on $e_7$ and $e_8$ can be undetected, its minimum distance is exactly 2.

Note that Fig. 1 as well as Freedman and Meye (2001, Fig. 2) is a surface code on a real projective plane, and that a projective plane is not homeomorphic to a spherical surface. In particular, edges $e_7$ and $e_8$ do not intersect with each other, and we have $f_3 = \{e_1, e_7, e_6, e_8\}$ and $f_4 = \{e_2, e_7, e_5, e_8\}$, which are impossible on a spherical surface. For details of surface codes on a projective plane, the readers are referred to Freedman and Meyer (2001).

### 6.1 Erasure at the seventh edge

In what follows, the edge $e_7$ is assumed to have an erased qubit. Then $I = \{e_7\}$, $\overline{I} = \{e_1, \ldots, e_6, e_8\}$, $\mathring{V}_I = \{v_2, v_4\}$, and $F_I = \{f_3, f_4\}$. We see that $D_X$ in Proposition 13 is spanned by $\{\vec{f_3}, \vec{f_4}\}$ and $D_Z$ is spanned by $\{\vec{v}_2, \vec{v}_4\}$. By Sect. 5.3, we see that $D_X$ can be spanned by only $\vec{f_3}$ and $D_Z$ can be spanned by only $\vec{v}_2$. Without our proposal, a decoder performs $7 = |\mathring{V}| + |F| - 2$ measurements on 8 qubits. With our proposal, 2 measurements are performed on 5 qubits at $\{e_1, e_5, e_6, e_7, e_8\}$. The numbers of measurements and measured qubits are much reduced by our proposal.

### 6.2 Erasure at the eighth edge

In what follows, the edge $e_8$ is assumed to have an erased qubit. By Sect. 5.3, we see that $D_X$ can be spanned by only $\vec{f_3}$ and $D_Z$ can be spanned by only $\vec{v}_1$. With our proposal, 2 measurements are performed on 6 qubits at $\{e_1, e_2, e_3, e_6, e_7, e_8\}$.

By a similar argument repeated with every edge, we also see that any single erasure can be corrected by measuring at most 6 qubits, and that the surface code in Fig. 1 has locality 5, which exemplifies Proposition 18.

## 7 Concluding remarks

In this paper, in order to correct erasures, we clarified which measurements are sufficient on which qudits in a stabilizer codeword with erasures in Theorem 3 and Proposition 4. Then we clarified the minimum required number of measurements by our proposal in Theorems 5 and 8, which looked similar to the dimension length profile of classical linear codes (Forney 1994). In Remark 6 we gave a computational procedure to find a smallest set of stabilizer observables to correct erasures by our proposal. Those general results for stabilizer codes were specialized to the CSS codes in Sect. 4.

As an application of the above results, in Sect. 5.2 localities of generalized surface codes were clarified in the context of quantum local recovery, and we also gave another computational procedure in Sect. 5.3 to find smallest sets of vertexes and faces for measurements to correct erasures, with a given generalized surface code and a given set $I$ of erasures. Consequently, we showed that correction of erasures at $I$ needs measurements of at most $|I|$ vertexes and at most $|I|$ faces. Unfortunately, it has cubic computational complexity $O(n^3)$ with code length $n$, as it is Gaussian elimination. Since erasure correction by using the full set of measurement outcomes on a generalized surface code can be done in linear computational complexity $O(n)$ (Delfosse and Zémor 2020), the cubic complexity is comparably large. A faster alternative to the procedure in Sect. 5.3 is strongly desired. Such an alternative may need to utilize structures specific to a class of codes for reduction of computational complexity, such as surface codes or LDPC codes, and further reduction of the complexity $O(n^3)$ seems difficult to this author for the general class of all stabilizer codes.

**Acknowledgements** This work is in part supported by the Japan Society for Promotion of Science under Grant No. 23K10980.

**Data availability** No datasets were generated or analyzed during the current study.

## Declarations

**Conflict of interest** The author has no conflict of interest to declare that are relevant to the content of this paper.